\documentclass[12pt]{article}
\usepackage[english]{babel}
\usepackage[utf8x]{inputenc}
\usepackage[T1]{fontenc}
\usepackage{physics}
\usepackage{wrapfig}
\usepackage{caption}
\usepackage{bbold}
\usepackage{amsmath}
\usepackage{graphicx}
\usepackage{graphicx, threeparttable}
\usepackage{xcolor}
\usepackage{enumitem}

\usepackage[a4paper,top=2.5cm,bottom=2.5cm,left=2.5cm,right=2.5cm,marginparwidth=2.5cm]{geometry}

\usepackage[colorinlistoftodos]{todonotes}
\usepackage[colorlinks=true, allcolors=blue]{hyperref}

\title{A Gödelian Hunch from Quantum Theory\footnote{FQXi Essay Contest 2020: https://fqxi.org/community/forum/topic/3518}}
\author{Hippolyte Dourdent\footnote{Univ. Grenoble Alpes, CNRS, Grenoble INP, Institut N\'eel, 38000 Grenoble, France ; hippolyte.lazourenko-dourdent@neel.cnrs.fr
}}

\date{\small{\today}}

\begin{document}

\maketitle
\section*{Introduction}

In classical logic, self-referring propositions can lead to pathologies such as the well-known Liar paradox ``This sentence is false.'' Because it features an over-determination - if the sentence is true then it is false, if it is false then it is true - the ``Liar'' leads to undecidability, the impossibility to decide whether the sentence is true or false. Analogs have been famously used in the foundations of mathematical logic, from Russell's paradox to Gödel's incompleteness theorem, passing by Tarski and Gödel undefinability theorem\footnote{The undefinability theorem stipulates that  any description of the truth of a proposition must be in a richer metalanguage than the language in which the proposition itself is stated ; this hierarchy of languages arising as a solution of the Liar.}. \\ 

In \cite{szangolies}, Szangolies coined the expression ``Gödelian hunch'' to describe ``the idea that the origin of the peculiarities surrounding quantum theory lie in phenomena related, or at least similar, to that of incompleteness in formal systems.'' \textit{What if the paradoxical nature of quantum theory could find its source in some undecidability analog to the one emerging from the Liar ?}  This essay aims at arguing for such quantum Gödelian hunch via two case studies: quantum contextuality as an instance of the Liar-like logical structure of quantum propositions ; and the measurement problem as a self-referential problem.\\

Quantum contextuality results from a theorem established by Kochen and Specker \cite{ks}, which shows that a quantum measurement cannot reveal a pre-existing value of a measured property independently of the measurement context. Using a narrative based on the Newcomb problem \cite{nozick}, the theological motivational origin of this result is introduced in order to show how the theorem might be related to a Liar-like undecidability (section \ref{sec:contextuality}). I will also briefly present a topological generalization of contextuality \cite{abramsky1} in which the logical structure of quantum contextuality is compared to sequences of cyclically referring statements, ``\textit{Liar cycles}'', which, associated with a truth predicate, lead to a logical contradiction \cite{abramsky3}. 

The measurement problem is often presented as a tension between the linear and deterministic evolution of the wave-function following the Schrödinger equation and the  projection postulate. Nevertheless, the problem was also analyzed as emerging from a logical error, and occurs because no distinction is made between theoretical and meta-theoretical objects. I will present my analysis of the related Wigner's friend thought experiment \cite{wigner} and a recent paradox by Frauchiger and Renner \cite{frauchiger}, introducing the notion of ``meta-contextuality'' as a Liar-like feature underlying the neo-Copenhagen interpretations of quantum theory (section \ref{sec:meas}).

Finally, this quantum Gödelian hunch opens a discussion of the paradoxical nature of quantum physics (section \ref{sec:para}) and the emergence of time itself from self-contradiction (section \ref{sec:time}).

\section{A Gödelian Hunch from Quantum Contextuality}
\label{sec:contextuality}

In 1960, inspired by Birkhoff and von Neumann's axiomatic approach to derive quantum theory from non-classical ``experimental propositions'' adapted to the experimental result of quantum mechanics, Specker asked:  `` \textit{Is it possible to extend the description of a quantum mechanical system through the introduction of supplementary - fictitious - propositions in such a way that in the extended domain the classical propositional logic holds [...] ?}'' \cite{specker,seevinck} The answer is negative, ``except in the case of Hilbert spaces of dimension 1 and 2.'' A fruitful collaboration with Kochen will culminate in an enriched reformulation of this result, today known as the Kochen-Specker theorem \cite{ks}. Thus, either a measurement reveals a pre-existing value of a measured property depending on the measurement context (quantum contextuality), or such value is unpredictable\footnote{For example the outcome might be brought-into-being by the act of measurement itself, ``Unperformed measurements have no results.''\cite{peres}} \cite{abbott}.

\subsection{Counterfactual Undecidability}
\label{counterfactual}
In his seminal work,  Specker noticed an analogy between these simultaneously undecidable propositions of quantum theory and the undecidability of \textit{counterfactual} propositions\footnote{A counterfactual proposition is a special kind of conditional proposition which follows the structure: ``If $A'$ had happened instead of $A$, then $B'$ would have happened instead of $B$.''}. Hence, the question of an extension of quantum propositions in classical logic is paralleled with: ``\textit{the scholastic speculations about the ``Infuturabilien " [...], that is, the question whether the omniscience of God also extends to events that would have occurred in case something would have happened that did not happen. (cf. e.g. [3], Vol. 3, p.363.)}" \cite{seevinck}. Can an omniscience extend to counterfactual propositions ? A possible positive answer is given by the reference ``([3], Vol. 3, p.363)''. The latter leads to a chapter on \textit{molinism}, an unorthodox form of omniscience proposed by scholastics in order to conciliate God's foreknowledge and human's free will. According to this view, if God had predicted that you will make a certain choice $A$, it may nevertheless have been in your power to do something, such that were you to do it, God would not have predicted this peculiar choice $A$. In a sense, God's omniscience and human free will can co-exist at the condition that the former is \textit{contextualized} by the latter.\\

In order to illustrate the afored mentioned analogy, I propose the following narrative. 
We invoke an omniscient demon whose omniscience extends to counterfactual propositions. Two observables $A$ and $B$ are given to a free agent, Alice. Alice can choose to measure the observable $B$ in two contexts: $C_1:=(A,B)$ or $C_2:=(B)$. Beforehand, the demon has predicted her choice and, based on it, has assigned a value to $B$: $v(B)_{|C_1}=0$ or $v(B)_{|C_2}=1$. Alice measures the value of the observable in one of the contexts, and assume that she verifies that the demon's prediction is correct. One can then ask the counterfactual question: what would have happened if she had chosen the other context? Two solutions are possible:
\begin{itemize}
    \item (a) ``If Alice had chosen the other context, she would have found a different value for $B$.'' In this case, the omniscience of the demon may extend to counterfactuals. But this implies that either Alice is not free of her choices (superdeterminism), or the omniscience of the demon is conditioned by the context she chooses (molinism). 
    \item (b) ``If Alice had chosen the other context, she would have found the same value for $B$.'' In this case, the omniscience of the demon does not extend to counterfactuals. The demon would have been wrong. Because its essence is defined by its function, denying it is an exorcism. Thus, the value of $B$ is \textit{unpredictable}. 
\end{itemize}

This narrative is freely inspired by the Newcomb problem \cite{nozick}, a decision theory problem where the values associated to the prediction correspond to distinct earnings (e.g. $v(A)=10k\$$, $v(B)_{|C_1}=1k\$$ and $v(B)_{|C_2}=1M\$$), the problem arising from the question of which choice allows Alice to maximize her gains. 
Interestingly, the problem might originate from a self-referential structure: ``\textit{Newcomb's problem may be understood as a game against one's self in which one's choice is based on deliberations that attempt to incorporate the outcome of this very choice.}'' \cite{slezak}\\

A similar ``circularity'' lies under the counterfactual statements (a) and (b).  It is of course trivial to point out that nothing is quantum in the Newcomb narrative. Yet, the non-Boolean logical structure of quantum theory yields analog conclusions: either a value-assignement to all observables is contextual or one cannot assign predefined values to all observables, i.e. these values are in general unpredictable.  The self-referential nature of these narratives hints at the presence of a similar circular structure underlying quantum contextuality. Approaching contextuality as the fact that quantum theory is based on intertwined Boolean algebras that cannot be embedded in a global Boolean algebra highlights this Liar-like structure.

\subsection{Topological Undecidability}

\begin{figure}[ht!]
\centering
\includegraphics[width=1\textwidth]{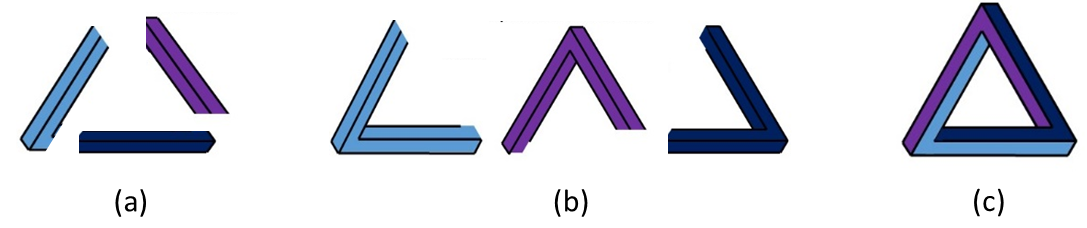}
\caption{(a) Each individual bar represents an observable to which one assign a truth-value. (b) Each observable is compatible with the other two separately, and thus two local contexts can be defined per observable. The truth values assigned to observables in a context are logically consistent. (c) The global picture of all bars glued together is an undecidable figure, the Penrose triangle. One cannot define a global context in which no truth-value assignment leads to a contradiction.}
\label{fig:penrose}
\end{figure}

In a topological approach by  Abramsky et al.\cite{abramsky1} based on sheaf theory and cohomology, contextuality emerges when data which are locally consistent are globally inconsistent. One can illustrate this definition by an analogy with famous undecidable figures such as the Penrose triangle\footnote{This is a figurative illustration which has a didactic purpose. Of course, sheaf-theoretic contextuality cannot be reduce to this simple example. Moreover, the Penrose triangle cannot be paralleled with quantum contextuality, where no proof of the Kochen-Specker theorem can be made out of three observables.} (Fig.\ref{fig:penrose}).

As noticed in \cite{abramsky3}, there is a direct connection between contextuality and classical semantic paradoxes called ``Liar cycles'', defined as sequences of statements of the form:
$[\{S_1,S_2\}$ true  ; ... ; $\{S_{N-1},S_N\}$ true ; $\{S_{N},S_1\}$ false] with $S_i$ the $i^{th}$ assertion, and $\{S_{i-1},S_i\}$ and $\{S_i,S_{i+1}\}$ the two ``local" contexts of this assertion. Although every proof of the Kochen-Specker theorem features such logical global obstruction, this generalized approach does not reduce to quantum contextuality, and also incorporates non-locality as a special case. As an example, the Hardy paradox \cite{hardy0} can be shown to entail contextuality, and thus feature a Liar-like logical structure (cf. details in appendix).\\

In such contextuality scenario, the contradiction occurs at the level of classical statements, inferred from quantum propositions. The assigned values are both classical and meta-theoretical, in the sense that they are not part of quantum theory.  Hence, if meta-theoretical statements are attached to quantum propositions, they cannot be embedded in a global Boolean one in general. The non-Boolean logic of quantum theory contaminates the meta-theoretical statements, which become globally \textit{undecidable}. I argue that this global undecidability of quantum propositions is in favor of a quantum Gödelian hunch.

\section{A Gödelian Hunch from the Measurement Problem} \label{sec:meas}

 As expressed in the literature, there exists different measurement problems (cf. e.g. \cite{brukner2}). The one we wish to tackle addresses `` the question of what makes a measurement a measurement. [...] There is nothing in the theory to tell us which device in the laboratory corresponds to a unitary transformation and which to a projection !'' \cite{brukner2}. This measurement problem as been analyzed as a ``logical error'' emerging from a lack of distinction between theoretical and meta-theoretical objects \cite{alexei2}. Similar conclusions explicitly underlying an analogy between the measurement problem and Gödel's theorem have been made (cf. \cite{szangolies} for an overview). For example, Chiara notices that such analysis could seem ``to be very close to some similar limitative results that we have accepted in logic such as the Gödel theorem (who realizes a proof of the consistency of a well-behaved scientific theory, must be `external' with respect to the theory (in the sense that he cannot use only the proof theoretical tools allowed by the theory)) [...].'' \cite{chiara}  I will analyze the Wigner's Friend thought experiment and  the Frauchiger-Renner paradox -which shows that ``a self-referential use of quantum theory yields contradictory claims.'' \cite{frauchiger} - as sustaining this Gödelian hunch.

\subsection{Wigner's Friend, Universality, Meta-Contextuality and Measurement }

 The measurement problem we are dealing with is usually formalized as follow. A quantum system is in the state $\ket{\psi}=\alpha\ket{0}+\beta\ket{1} \in\mathcal{H_S}$. On the one hand, following the projection postulate, the system will either be projected onto state $\ket{0}$ with probability $|\alpha|^2$, or state $\ket{1}$ with probability $|\beta|^2$ after the measurement. On the other hand, if the ``observer'' (e.g. the measuring device) is a physical system, then it shall be described by quantum theory. A Hilbert space $\mathcal{H_O}$ is associated to this observing system. Defining $\ket{M}$ the observer
 state ``ready to perform a measurement'', the initial state of the compound system in $\mathcal{H}_S\otimes\mathcal{H}_O$ is $(\alpha\ket{0}+\beta\ket{1})\otimes\ket{M}$. In this case, the measurement process is described as an interaction between the system and the device, and thus as a unitary transformation $U$, resulting in
$U[(\alpha\ket{0}+\beta\ket{1})\otimes\ket{R}]\rightarrow \alpha\ket{0}\otimes\ket{M_0}+\beta\ket{1}\otimes\ket{M_1} $. Because the two final states are physically distinct, there seem to be a tension between the postulates of quantum theory, raising the question of how a measurement process should be described.\\

The Wigner's Friend thought experiment \cite{wigner} is a meta-illustration of this measurement problem, which asks: what happens when an observer observes another observer observing a quantum system ? A quantum system, e.g. a qubit living in $\mathcal{H}_S$, is given to an observer, Wigner's friend, who can perform a measurement on this system in her laboratory. Outside her laboratory, another observer, Wigner, can associate a quantum state to the compound system  $\mathcal{H}_S\otimes\mathcal{H}_O$, where $\mathcal{H}_O$ is a Hilbert space associated to Wigner's friend, e.g. a memory qubit $\ket{M_i}$ who can be interpreted as ``Wigner's friend observes a projection on state $\ket{i}$''. While Wigner's friend observes a collapse of the qubit, the measurement process has been described as a unitary transformation from Wigner's perspective. However both descriptions should be valid.\\

My analysis of this problem relies on the following terminology. The quantum system is an \textit{object}, since it is described by quantum theory. Wigner's friend is an observer, and as a user of quantum theory, is a meta-theoretical object, in short a \textit{meta-object}. Wigner is an observer who can perform a measurement on systems of the form object $\otimes$ meta-object, and is thus a meta-meta-object, or \textit{meta-observer}. The problem seems to arise from the fact that an observer and a meta-observer are lead to describe the same event in contradictory ways.
I introduce the notion of \textit{meta-context} as a set of the form \{meta-object,object\}. This set is defined by a movable \textit{cut} between theoretical objects studied in the language of the theory and meta-theoretical objects which are out of the range of the theory. In the Wigner's friend paradox, two meta-context  are involved:  \{Wigner's friend, $\mathcal{H_S}$\} and \{Wigner, $\mathcal{H_S}\otimes\mathcal{H_O}$\}.\\

\begin{figure}[ht!]
\centering
\includegraphics[width=1\textwidth]{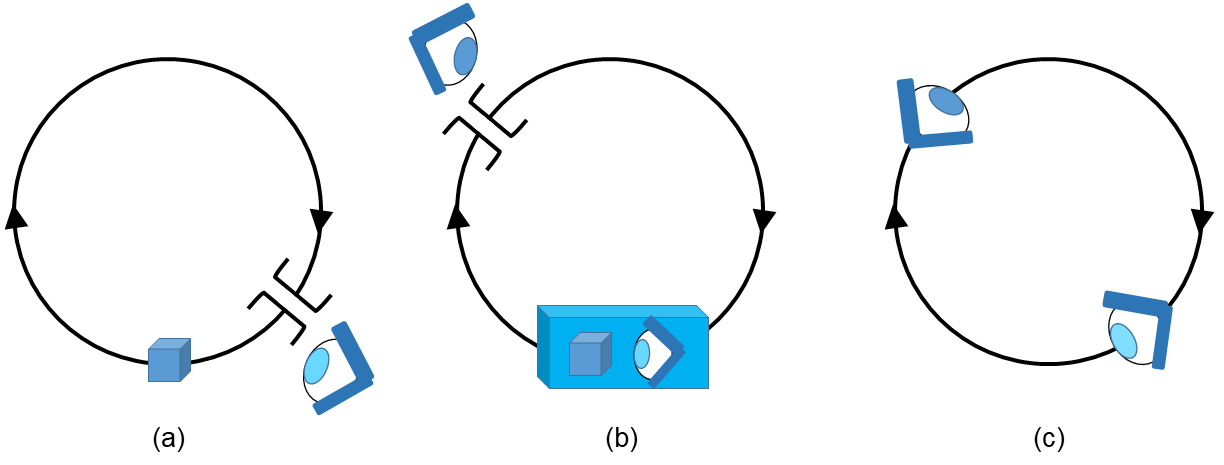}
\caption{Inspired by Grinbaum's epistemic loops \cite{alexei2}, let us represent all theoretical objects by a loop. Cutting the loop sends objects in the meta-theoretical domain. (a) Meta-Context \{Wigner's friend, $\mathcal{H_S}$\}.  (b) Meta Context \{Wigner, $\mathcal{H_S}\otimes\mathcal{H_O}$\}. (c) Maintaining (Q) and (NMC) leads to ignoring the relative cuts, i.e. the meta-contexts. Wigner and Wigner's friend are put at the same level, and self-referential inconsistencies may occur. }
\label{wigner}
\end{figure}

The problem can be understood as follows. Firstly, quantum theory is assumed to be correct and can be applied to any object whatsoever. Such assumption is called \textbf{\textit{quantum universality}} (Q). Secondly, one assumes that truth values given by the propositions associated with an object are independent of the meta-context, of whether the object is theoretical or meta-theoretical, i.e. the truth values are \textbf{\textit{non-meta-contextual}} (NMC)\footnote{This notion is equivalent to Brukner's ``observer-independent facts'' \cite{brukner1}.}. Maintaining (Q) and (NMC) leads to an absolute form of universality: everything can be described by the theory, irrespective of the meta-context, no cut is needed. But imagine an infinite chain of observers observing observers observing a quantum system. Then,  meta- ... -meta-observers are invoked, \textit{ad infinitum}. One could argue that the ultimate meta$^\infty$-object is God, or some Laplacian demon. However, if such a demon can measure the whole Universe, then the demon is necessarily excluded from the Universe in order to avoid Liar-like inconsistencies, independently of the considered theory. As shown by Breuer \cite{breuer1}, if a theory is considered to be absolutely universally valid, then the theory cannot be experimentally fully accessible, due to self-referential problems. There is a tension between absolute universality (Q,NMC), in which the measuring process might be treated theoretically, and \textbf{\textit{measurement} } as a meta-theoretical process. In the light of this analysis, the most appealing solution is to drop (NMC) and acknowledge the observer for what it is: a meta-object. This way, the notion of meta-observer becomes obsolete, and the logical inconsistencies are avoided (cf. Figure \ref{wigner}). The universality of the theory is maintained, but becomes \textit{relative}. Any object can be cut and become a meta-object. However, once the cut is fixed, any out-of-meta-context question is \textbf{\textit{undecidable}}. ``Although it can describe \textit{anything}, a quantum description cannot include \textit{everything}.''\cite{peres}

\subsection{``Wigner's Friendifications''}

Recently, there has been a renewed interest in Wigner's thought experiment in the field of quantum foundations. This resurgence is due to the appearance of new hybrid paradoxes \cite{brukner1,frauchiger}, which rely on a Wigner's Friendification\footnote{In my knowledge, this terminology was first used by Aaronson in a blog post (www.scottaaronson.com/blog/?p=3975) in order to describe the Frauchiger-Renner paradox. }, a transformation of previous quantum ``paradoxes'' where one allows meta-objects to be described as objects of the theory, and allows meta-observers to measure coumpound systems of the type ``object $\otimes$ meta-object''. I will analyze the Frauchiger-Renner paradox \cite{frauchiger} as a Wigner's Friendification of the Hardy paradox explicitly showing the logical inconsistency which can emerge from (Q,NMC).\\

The original Hardy scenario involved two agents/observers, Alice and Bob, sharing a two-qubit system. In the new thought experiment, Alice and Bob are upgraded to meta-observers, while two new agents, their respective friends, share a two-qubit system and can perform a measurement on their respective part of the system. Like in the standard scenario, Alice and Bob's friend can measure their qubit in the computational $\{\ket{0} , \ket{1} \}$ basis or in the diagonal $\{\ket{+} , \ket{-} \}$ basis. Regarding Alice and Bob, these bases are ``Wigner's friendified'' as follows. The computational basis is transformed into a meta-computational basis corresponding to an ``observer basis'', a statement made by the observer, the friend\footnote{ More precisely, it corresponds to a meta-observer asking her friend in which state has the qubit been projected.}: $\{\ket{0}_{S_A}\otimes\ket{0}_{F_A},\ket{1}_{S_A}\otimes\ket{1}_{F_A}\}$. For example, if Alice's friend finds his qubit in state $\ket{0}_{S_A}$, then his statement will be $\ket{0}_{F_A}$ and Alice will find the global system in the state $\ket{0}_{S_A}\otimes\ket{0}_{F_A}$. The diagonal basis of the standard observation becomes a meta-diagonal basis corresponding to a ``meta-observer basis'', where the meta-observer actually performs a quantum measurement on the compound system, resulting in a statement associated to the meta-observer:$\{\ket{+}_A,\ket{-}_A\}$, with $\ket{\pm}_A=\frac{1}{\sqrt{2}}(\ket{0}_{S_A}\otimes\ket{0}_{F_A}\pm\ket{1}_{S_A}\otimes\ket{1}_{F_A})$. Applying this Wigner's Friendification to the four sentences of the Hardy paradox (cf appendix), one obtains four new assertions:  \\

\textbf{Sentence FR1}: `` ``If Alice obtains ``$-$'', then Bob's friend obtains outcome ``$1$''.''\\ 

\textbf{Sentence FR2}: ``If Bob's friend obtains ``$1$'', then Alice's friend obtains outcome ``1''.'' \\

\textbf{Sentence FR3}: `` ``If Alice's friend obtains ``$1$'', then Bob obtains outcome ``$+$''.''\\ 

\textbf{Sentence FR4}: ``Alice and Bob both obtain ``$-$'' with a probability of $\frac{1}{12}$.'' \\

Like in the Hardy paradox, these sentences forms a probabilistic Liar cycle: assume that Bob and Alice both obtains `$-$' (this happens with a probability 1/12). Bob obtains ``$-$'' and Alice obtains `$-$'' $\rightarrow$ Bob's friend obtains ``$1$'' $\rightarrow$ Alice's friend obtains ``$1$'' $\rightarrow$ Bob obtains ``$+$'', contradicting the first statement. In \cite{frauchiger}, the authors analyze this paradox as an incompatibility between three assumptions: (Q) quantum theory is correct and can be applied to systems of any complexity ; (C) observers and meta-observers claims should be consistent with each other ; (S) a measurement yields a single outcome.
Assumption (C), in particular,  has been widely discussed in the literature (cf. for example \cite{brukner1,bub,fortin}). I argue that this assumption can be reformulated into two assumptions: non-contextuality and non-meta-contextuality.\\

Indeed, like the Hardy paradox, the Frauchiger-Renner paradox entails contextuality in the sense of Abramsky: a global logical obstruction of four consistent propositions\footnote{ Note that the paradox has already been analyzed as applying classical logic to quantum propositions which is forbidden by the non-Boolean structure of quantum theory \cite{brukner1,bub,fortin}}. Thus the contradiction might occur from assuming non-contextuality (NC). However, unlike the Hardy paradox, here each statement can be associated to one agent: one for each observer (FR2 and FR3), and one for each meta-observer (FR1 and FR4). In fact, like in the original Wigner's friend experiment meta-objects (the friends) are described in the language of the theory, i.e. at the level of objects. As seen previously, this is equivalent to the (NMC) assumption, which associated with (Q), can lead to self-referential inconsistencies when statements made in different meta-contexts are compared. Giving up on (NMC), consistency is restored, but only inside a meta-context among \{Alice, Alice's Friend $\otimes$ qubit $S_A$ \} ; \{Bob, Bob's Friend $\otimes$ qubit $S_B$\} ; \{Alice's Friend, qubit $S_A$ \} ; \{Bob's Friend, qubit $S_B$ \}.  Under such analysis, the fact that ``a self-referential use of the theory yields contradictory claims'' \cite{frauchiger} is not especially surprising, if one acknowledge that quantum theory can only be consistently used in a meta-context, i.e. that \textbf{\textit{the use of quantum theory is (meta-)contextual}}. 

\subsection{The Heirs of Copenhagen}

Analyzing the measurement problem as self-reference and escaping the logical inconsistency by introducing a \textit{cut}\footnote{Sometimes called the von Neumann or Heisenberg cut (``Schnitt'').} complies with various ``neo-Copenhagen'' interpretations of quantum theory, often wrongly labeled as ``anti-realistic'' \cite{fuchs2}, such as information-based interpretations \cite{bub1,brukner3,rovelli} and QBism \cite{fuchs3}. All agree on the fundamental distinction between the meta-theoretical and theoretical object. In these interpretations, this movable cut is \textit{functional} and not ontological. It does not discriminate a macroscopic classical world from a microscopic quantum one, because every object can be treated by the theory (Q) or not. This is especially made explicit in Rovelli's relational interpretation \cite{rovelli}. Following the footsteps of Bohr: ``There is no quantum world. There is only an abstract quantum physical description. It is wrong to think that the task of physics is to find out how nature \textit{is}. Physics concerns what we can \textit{say} about nature. We depend on our words, our task is to communicate experience and ideas to others. We are suspended in language ...'' \cite{petersen} ; or as Wittgenstein wrote in his \textit{Tractatus}: ``(5.632) The subject does not belong to the world: rather it is a limit of the world.''  Absolute universality has a God-like flavour and leads to paradoxical features that cannot be \textit{said}. On the contrary, one can acknowledge the transcendental status of the meta-theoretical object: a classical (Boolean) description is the condition of possibility for the rendering of quantum (non-Boolean) events.

\section{Conclusion: Is Physics Paradoxical ?} \label{sec:para}

In his seminal paper on the logic of simultaneously undecidable propositions \cite{specker, seevinck}, Specker attached the following epigraph: ``La logique est d'abord une science naturelle.'' [Logic is in the first place a natural science.] extract from ``La physique de l'objet quelconque'' by Gonseth. Gonseth argued that logic should be considered as an experimentally refutable science of  ``any object whatsoever''. If quantum physics goes against classical logic, thus classical logic should be revised. Several years later, Putnam defended a similar idea in a paper entitled `Is Logic Empirical ?'' \cite{putnam}. Mirroring this interrogation, we ask: ``Is Physics Paradoxical ?''.\\

Quantum theory does not only defy common sense, but it also defies classical logic, i.e. our common language and semantic. In this sense, quantum theory is more paradoxical than other physical theories. But is Nature itself paradoxical ?  Does the world really feature intrinsically strange phenomena that cannot be grasped with our words, whether it is a non-local behaviour or parallel worlds ? In this essay, I argued for an alternative. Quantum paradoxes are not physical, but emerge from \textbf{\textit{a lack of metaphysical distancing}}.
I highlighted how the Liar-like structure of quantum propositions enlightened by the Kochen-Specker theorem already invites to give up on considering quantum objects as entities with intrinsic properties independently of the questions asked by a meta-theoretical object. I proposed the notion of ``meta-contextuality'' to explain how neo-Copenhagen interpretations avoid the measurement problem, Wigner's friend and Wigner's friendified paradoxes by analyzing them as logical errors. Acknowledging the need for an undiscriminating cut between meta-theoretical and theoretical objects when one uses quantum theory, any question that ignores this transcendental distinction looses its operational significance and becomes physically undecidable. Thus, quantum paradoxes might just be instances of a fundamental undecidability, contributing to a quantum Gödelian hunch\footnote{A very recent result \cite{ji} also contributes to the quantum Gödelian hunch. Using a modified proof of quantum contextuality, the authors proved that the class MIP* of problems that can be decided by a polynomial-time referee interacting with quantum agents sharing entanglement contains Liar-like undecidable problems.}. Finally, this essay fully adheres to Wheeler's intuition\footnote{Wheeler  might have been one of the first to investigate this quantum Gödelian hunch. A famous anecdote tells that Wheeler was thrown out of Gödel’s office for asking him if there was a connection between his incompleteness theorem and  Heisenberg’s uncertainty principle. \cite{szangolies}}: ``Physics is not machinery. Logic is not oil occasionally applied to that machinery. Instead, everything, physics included, derives from two parents, and is nothing but cathode-tube image of the interplay between them. One is the ``participant''. The other is the complex of undecidable propositions of mathematical logic.'' \cite{fuchs2}

\section{Epilogue: A Gödelian Hunch from Time} \label{sec:time}

In 1949,  Gödel discovered solutions of general relativity later known as closed time-like curves (CTCs) which theoretically would allow an observer to travel back in her own past \cite{godel}. However, the existence of such closed causal loops seems to imply the possibility for a traveller to interact with her own past-self, and for example prevent her own time-travel. This paradox, known as the grandfather antinomy, shares the same logical structure as the Liar. Unlike quantum theory, where the Gödelian hunch relies on the semantic of the theory, the grandfather paradox is a (speculative) \textit{physical realization} of a self-contradiction.\\

By analogy with the scholastic debate previously introduced in \ref{counterfactual}, the paradox can be understood as the tension between events that already happened and the ability to decide whether these ``physically-already-determined'' facts can be changed or not. Here, the role of God or the omniscient demon is played by \textit{time} itself. Thus, one could deny time its fundamental aura, and argue instead that it is emergent. In fact, inside a closed loop, ``time'' is undefinable. Nevertheless, facing a global inconsistency, one can cut the loop, and recover logical consistency. As Gödel wrote: ``\textit{Time is the means by which God realized the inconceivable that P and non-P are both true} [...].'' \cite{cassou} This way, time emerges from cutting self-referential paradoxes. Noticing that this cut might be epistemic, in line with a Gödelian hunch, one could finally speculate that ``\textit{Time is a consequence of every attempt to provide a comprehensive description of the universe from within. Thus, time in this sense is not related to the universe itself but to the attempt to describe it.}’’\cite{kull}

\subsection*{Acknowledgments}

I would like to thanks Cyril Branciard for his precious support and advices, and Alexei Grinbaum for inspiring discussions.

\newpage
\bibliographystyle{ieeetr}
\bibliography{sampleb}

\begin{thebibliography}{10}

\bibitem{szangolies}
J.~{Szangolies}, ``{Epistemic Horizons and the Foundations of Quantum
  Mechanics},'' {\em ArXiv e-prints}, May 2018.

\bibitem{ks}
S.~Kochen and E.~Specker, ``{T}he {P}roblem of {H}idden {V}ariables in
  {Q}uantum {M}echanics,'' {\em Journal of Mathematics and Mechanics}, vol.~17,
  no.~1, pp.~59--87, 1967.

\bibitem{nozick}
R.~Nozick, ``Newcomb's problem and two principles of choice,'' in {\em Essays
  in Honor of Carl G. Hempel} (N.~Rescher, ed.), pp.~114--146, Reidel, 1969.

\bibitem{abramsky1}
S.~Abramsky, S.~Mansfield, and R.~S. Barbosa, ``The cohomology of non-locality
  and contextuality,'' {\em arXiv preprint arXiv:1111.3620}, 2011.

\bibitem{abramsky3}
S.~Abramsky, R.~S. Barbosa, K.~Kishida, R.~Lal, and S.~Mansfield,
  ``Contextuality, cohomology and paradox,'' {\em arXiv preprint
  arXiv:1502.03097}, 2015.

\bibitem{wigner}
E.~P. Wigner, ``{Remarks on the Mind-Body Question},'' in {\em The Scientist
  Speculates} (I.~J. Good, ed.), Heineman, 1961.

\bibitem{frauchiger}
D.~Frauchiger and R.~Renner, ``{Quantum theory cannot consistently describe the
  use of itself},'' {\em Nature Communications}, vol.~9, p.~3711, 2018.

\bibitem{specker}
E.~Specker, ``Die {L}ogik {N}icht {G}leichzeitig {E}ntscheidbarer {A}ussagen,''
  {\em Dialectica}, vol.~14, no.~2/3, pp.~239--246, 1960.

\bibitem{seevinck}
M.~P. Seevinck, ``E. specker:" the logic of non-simultaneously decidable
  propositions"(1960),'' {\em arXiv preprint arXiv:1103.4537}, 2011.

\bibitem{peres}
A.~Peres and W.~H. Zurek, ``Is quantum theory universally valid?,'' {\em
  American Journal of Physics}, vol.~50, no.~9, pp.~807--810, 1982.

\bibitem{abbott}
A.~A. Abbott, C.~S. Calude, and K.~Svozil, {\em On the Unpredictability of
  Individual Quantum Measurement Outcomes}, pp.~69--86.
\newblock Cham: Springer International Publishing, 2015.

\bibitem{slezak}
P.~Slezak, ``Demons, deceivers and liars: Newcomb's malin g{\'e}nie,'' {\em
  Theory and Decision}, vol.~61, pp.~277--303, Nov 2006.

\bibitem{hardy0}
L.~Hardy, ``Nonlocality for two particles without inequalities for almost all
  entangled states,'' {\em Phys. Rev. Lett.}, vol.~71, pp.~1665--1668, Sep
  1993.

\bibitem{brukner2}
{\v{C}}.~Brukner, {\em On the Quantum Measurement Problem}, pp.~95--117.
\newblock Cham: Springer International Publishing, 2017.

\bibitem{alexei2}
A.~Grinbaum, ``On epistemological modesty,'' {\em Philosophica}, vol.~83,
  pp.~139--150, 2010.

\bibitem{chiara}
M.~Chiara, ``{Logical Self Reference, Set Theoretical Paradoxes and the
  Measurement Problem in Quantum Mechanics},'' {\em Journal of Philosophical
  Logic}, vol.~6, no.~1, pp.~331--347, 1977.

\bibitem{brukner1}
{\v C}.~Brukner, ``A no-go theorem for observer-independent facts,'' {\em
  Entropy}, vol.~20, no.~5, 2018.

\bibitem{breuer1}
T.~Breuer, {\em John von Neumann Met Kurt G{\"o}del: Undecidable Statements in
  Quantum Mechanics}, pp.~159--170.
\newblock Dordrecht: Springer Netherlands, 1999.

\bibitem{bub}
J.~Bub, ``In defense of a “single-world” interpretation of quantum
  mechanics,'' {\em Studies in History and Philosophy of Science Part B:
  Studies in History and Philosophy of Modern Physics}, 2018.

\bibitem{fortin}
S.~Fortin and O.~Lombardi, ``Wigner and his many friends: A new no-go
  result?,'' 2019.

\bibitem{fuchs2}
C.~A. Fuchs, ``On participatory realism,'' in {\em Information and
  interaction}, pp.~113--134, Springer, 2017.

\bibitem{bub1}
J.~Bub and I.~Pitowsky, ``Two dogmas about quantum mechanics,'' {\em Many
  worlds}, pp.~433--459, 2010.

\bibitem{brukner3}
{\v{C}}.~Brukner and A.~Zeilinger, ``Quantum physics as a science of
  information,'' in {\em Quo Vadis Quantum Mechanics?}, pp.~47--61, Springer,
  2005.

\bibitem{rovelli}
C.~Rovelli, ``Relational quantum mechanics,'' {\em International Journal of
  Theoretical Physics}, vol.~35, pp.~1637--1678, Aug 1996.

\bibitem{fuchs3}
C.~A. Fuchs and B.~C. Stacey, ``Qbism: Quantum theory as a hero’s handbook,''
  in {\em Proceedings of the International School of Physics “Enrico Fermi},
  vol.~197, pp.~133--202, 2019.

\bibitem{petersen}
A.~Petersen, ``{The Philosophy of Niels Bohr},'' {\em Bulletin of the Atomic
  Scientists}, vol.~19, no.~7, pp.~8--14, 1963.

\bibitem{putnam}
H.~Putnam, ``Is logic empirical?,'' in {\em Boston studies in the philosophy of
  science}, pp.~216--241, Springer, 1969.

\bibitem{ji}
Z.~Ji, A.~Natarajan, T.~Vidick, J.~Wright, and H.~Yuen, ``Mip*= re,'' {\em
  arXiv preprint arXiv:2001.04383}, 2020.

\bibitem{godel}
K.~G\"odel, ``{An Example of a New Type of Cosmological Solutions of Einstein's
  Field Equations of Gravitation},'' {\em Rev. Mod. Phys.}, vol.~21,
  pp.~447--450, Jul 1949.

\bibitem{cassou}
P.~Cassou-Nogu{\`e}s, {\em Les D{\'e}mons de G{\"o}del. Logique et folie:
  Logique et folie}.
\newblock Le Seuil, 2015.

\bibitem{kull}
A.~Kull, {\em Self-Reference and Time According to Spencer-Brown}, pp.~71--79.
\newblock Berlin, Heidelberg: Springer Berlin Heidelberg, 1997.

\end{thebibliography}

\newpage
\normalsize
\section*{Technical Appendices}

\subsection*{The Hardy Paradox}

In this scenario, two agents, Alice and Bob, share  a two-qubits system in a specific entangled state. Each agent can choose to measure their respective qubit in a computational $\{\ket{0},\ket{1}\}$ or a diagonal basis $\{\ket{+},\ket{-}\}$ with $\ket{\pm}=\frac{1}{\sqrt{2}}(\ket{0}\pm\ket{1})$. The initial entangled state can thus be written in four different basis, each corresponding to a measurement context. For example, in the comput.-comput. basis, the state is: $\ket{\psi}=\frac{1}{\sqrt{3}}(\ket{00}+\ket{10}+\ket{11})$. Assuming that a predefined value can be associated to a measured property when a result can be predicted with certainty, one can infer the four following sentences, each associated to a measurement context:\\

(1) $\bullet$  In the diago.-comput. basis, the state before measurements is:\\

$\ket{\psi}=\sqrt{\frac{2}{3}}\ket{+0}+\frac{1}{\sqrt{6}}\ket{+1}-\frac{1}{\sqrt{6}}\ket{-1}$\\

\textbf{Sentence H1} : ``If Alice obtains `$-$', then Bob obtains `$1$'.''\\

(2) $\bullet$ In the comput.-comput.  basis, the state before measurements is:\\

$\ket{\psi}=\frac{1}{\sqrt{3}}(\ket{00}+\ket{10}+\ket{11})$\\

\textbf{Sentence H2} : ``If Bob obtains `$1$', then Alice obtains `$1$'.''\\

(3) $\bullet$ In the comput.-diago.  basis, the state before measurements is:\\

$\ket{\psi}=\sqrt{\frac{2}{3}}\ket{1+}+\frac{1}{\sqrt{6}}\ket{0+}+\frac{1}{\sqrt{6}}\ket{0-}$\\

\textbf{Sentence H3} : ``If Alice obtains `$1$', then Bob obtains `$+$'.''\\

(4) $\bullet$ In the diago.-diago.  basis, the state before measurements is:\\

$\ket{\psi}=\frac{3}{\sqrt{12}}\ket{++}+\frac{1}{\sqrt{12}}\ket{+-}-\frac{1}{\sqrt{12}}\ket{-+}+\frac{1}{\sqrt{12}}\ket{--}$\\

\textbf{Sentence H4} : ``Alice and Bob can both obtain `$-$' with a probability 1/12.''\\

Assuming non-contextuality means that one can build inferences from these different sentences. For instance, from $(H1, H2, H3)$, one can construct the sentence: ``If Alice obtains `$-$', then Bob obtains `$+$' ''. However, this sentence is incompatible with $H4$. Thus, $((H1,H2,H3),H4)$ is globally inconsistent, and the paradox entails contextuality. The following probabilistic\footnote{The Hardy paradox is a probabilistic Liar cycle because the contradiction only occurs with a probability 1/12.} Liar cycle can be formulated, assuming that both Alice and Bob obtained `$-$': Bob obtains `$-$' and Alice obtains `$-$' $\rightarrow$ Bob obtains `$1$' $\rightarrow$ Alice obtains `$1$' $\rightarrow$ Bob obtains `$+$', contradicting the first assignment.

\subsection*{``Wigner's Friendification'' of the Hardy Paradox}

The Hardy paradox presented above is Wigner's friendified as follows: The computational basis is transformed into a meta-computational basis corresponding to an ``observer basis''\\ $\{\ket{0}_{S_A}\otimes\ket{0}_{F_A},\ket{1}_{S_A}\otimes\ket{1}_{F_A}\}$. The diagonal basis of the standard observation becomes a meta-diagonal basis corresponding to a ``meta-observer basis'':$\{\ket{+}_A,\ket{-}_A\}$,\\ with $\ket{\pm}_A=\frac{1}{\sqrt{2}}(\ket{0}_{S_A}\otimes\ket{0}_{F_A}\pm\ket{1}_{S_A}\otimes\ket{1}_{F_A})$. The corresponding sentences can then be derived:\\

(1) $\bullet$  In the metaobserver-observer basis, the state before measurements is:
\[\ket{\psi}_{tot}=\sqrt{\frac{2}{3}}\ket{+}_A\ket{0}_{S_B}\ket{0}_{F_B}+\frac{1}{\sqrt{6}}\ket{+}_A\ket{1}_{S_B}\ket{1}_{F_B}-\frac{1}{\sqrt{6}}\ket{-}_A\ket{1}_{S_B}\ket{1}_{F_B}\]

\textbf{Sentence FR1}: ``If Alice finds the outcome `$-$', she knows that Bob's friend obtained outcome `$1$'.''\\

(2) $\bullet$  In the observer-observer basis, the state before measurements is:
\[\ket{\psi}_{tot}=\frac{1}{\sqrt{3}}\left(\ket{0}_{S_A}\ket{0}_{F_A}\ket{0}_{S_B}\ket{0}_{F_B}+\ket{1}_{S_A}\ket{1}_{F_A}\ket{0}_{S_B}\ket{0}_{F_B}+\ket{1}_{S_A}\ket{1}_{F_A}\ket{1}_{S_B}\ket{1}_{F_B}\right)\]

\textbf{Sentence FR2}: ``If Bob's friend finds the outcome `$1$', he knows that Alice's friend obtained outcome `1'.''\\

(3) $\bullet$  In the observer-metaobserver basis, the state before measurements is:
\[\ket{\psi}_{tot}=\sqrt{\frac{2}{3}}\ket{1}_{S_A}\ket{1}_{F_A}\ket{+}_B+\frac{1}{\sqrt{6}}\ket{0}_{S_A}\ket{0}_{F_A}\ket{+}_B+\frac{1}{\sqrt{6}}\ket{0}_{S_A}\ket{0}_{F_A}\ket{-}_B\]

\textbf{Sentence FR3}: ``If Alice's friend finds the outcome `$1$', she knows that Bob obtained outcome `$+$'.'' \\

(4) $\bullet$  In the metaobserver-metaobserver basis, the state before measurements is:
\[\ket{\psi}_{tot}={\frac{3}{\sqrt{12}}}\ket{+}_A\ket{+}_B+\frac{1}{\sqrt{12}}\ket{+}_A\ket{-}_B-\frac{1}{\sqrt{12}}\ket{-}_A\ket{+}_B+\frac{1}{\sqrt{12}}\ket{-}_A\ket{-}_B\]

\textbf{Sentence FR4}: ``Alice and Bob both find the outcome `$-$' with a probability of $\frac{1}{12}$.''\\

 The experiment is repeated $n$ times. A contradiction arises when the four statements are combined and when, for the nth round, Bob obtains outcome ``$-$'' and knows that Alice also obtains outcome ``$-$'' (FR4). From FR1, Bob knows then that Alice's friend obtained outcome ``$1$'', and thus, from FR2, that Bob's friend obtained outcome ``$1$''. But, from FR3, this implies that Bob knows that he himself obtained outcome ``$+$'', contradicting the fact that he obtained outcome ``$-$''.\\

\end{document}